# Consistent optical and electrical determination of carrier concentrations for the accurate modeling of the transport properties of *n*-type Ge

José Menéndez [1,a)], Chi Xu,[1] and John Kouvetakis [2]

[1]Department of Physics, Arizona State University, Tempe, AZ 85287-1504, USA
[2]School of Molecular Sciences, Arizona State University, Tempe, AZ 85287-1604, USA

a) Electronic mail: jose.menendez@asu.edu

Keywords: Germanium, doping, ellipsometry, Hall effect, mobility

**ABSTRACT**
A consistent methodology is presented to extract carrier concentrations in *n*-type Ge from measurements of the infrared dielectric function and the Hall effect. In the case of the optical measurements—usually carried out using spectroscopic ellipsometry—the carrier concentration is affected by the doping dependence of the conductivity effective mass, which is computed using a model of the electronic density of states that accounts for non-parabolicity and is fit to electronic structure calculations. Carrier concentrations obtained from Hall measurements require a knowledge of the Hall factor, which is arbitrarily set equal to unit in most practical applications. We have calculated the Hall factor for *n*-Ge using a model that accounts for scattering with phonons and with ionized impurities.
We show that determinations of the carrier concentration *n* using our computed effective mass and Hall factor virtually eliminates any systematic discrepancy between the two types of measurement. We then use these results to compute majority carrier mobilities from measured resistivity values, to compare with measurements of minority carrier mobilities, and to fit empirical expressions to the doping dependence of the mobilities that can be used to model Ge devices.

# I. INTRODUCTION

The 1950s and early 1960's represented the golden years of Ge technology. The development of the semiconductor transistor and the fabrication of detectors for nuclear applications led to dramatic improvements in Ge crystal growth techniques. The resulting material had an unprecedented purity, which allowed for very detailed studies of its basic properties. However, the subsequent introduction of MOS technology shifted the attention to Si, whose oxide lends itself almost ideally for device applications. In a few years, Si became the dominant semiconductor, and this role was reinforced by a concomitant progress in Si crystal growth, which lowered the cost of Si substrates dramatically. This made it increasingly difficult for other materials to compete, even in areas for which they have better properties than Si. In spite of these odds, germanium is currently enjoying a renaissance fueled by the replacement of $SiO_2$ by high-κ dielectrics and the development of viable epitaxial growth techniques that make it possible to deposit high-quality Ge films on Si-substrates, leveraging the progress in Si-technology.[1] New applications for Ge have emerged as well, in fields such as optoelectronics[2] and plasmonics.[3] Furthermore, the successful demonstration of $Ge_{1-y}Sn_y$ devices,[4] confirms that alloying with Sn is a viable tool for further fine-tuning the material properties. In particular, the alloy becomes a direct band gap semiconductor for modest Sn concentrations near 8%,(Ref. 5) which has led to the fabrication of group-IV lasers on Si substrates.[6]

The re-emergence of Ge as a prime semiconductor material requires the development of modern modeling tools based on a detailed knowledge of its physical properties. This ongoing effort has revealed significant gaps and some inconsistencies in the data collected in the early days of this technology. Most of these issues have to do with the properties of doped Ge. For example, in contrast to Si, very limited information is available on the difference in mobility between majority and minority electrons and holes. Even more basically, carrier concentrations have been determined almost universally from Hall effect measurements assuming a Hall factor equal to unity. In the case of electrons, the Hall carrier concentrations are matched by infrared reflectivity measurements if the conductivity effective mass is about 30% higher than the value determined from cyclotron resonance experiments.[7] Sixty years ago such discrepancy did not seem serious given the limited knowledge of the Ge band structure, but today it cannot be brushed aside in view of modern electronic structure calculations. Inconsistencies related to doping are exacerbated by the fact that modern applications require ultra-highly doped materials that were not widely available in the 50's and 60's. For these materials, degeneracy effects—including incomplete ionization—and electronic structure renormalization effects must be fully taken into account.

The introduction of novel precursors for low temperature *in situ* doping of Ge and GeSn layers, such as $SbD_3$, $As(GeH_3)_3$, $P(GeH_3)_3$, $As(SiH_3)_3$, and $P(SiH_3)_3$ (Refs. 8-12), have led to doped layers with extremely uniform doping profiles and very high levels of activation without annealing. These are ideal for fundamental studies, which have improved our knowledge of band gap renormalization effects,[13] revealed that incomplete ionization is virtually non-existent in Ge,[14] and enabled the observation of Fermi-level singularities in the dielectric function of *n*-type Ge.[15] Spectroscopic ellipsometry in the visible has played an important role in these experiments. The use of this technique in the infrared provides more information than reflectivity studies[16] because the real and imaginary part of the dielectric function are determined independently. In this paper, we combine the latest experimental results and a theoretical analysis of Hall and ellipsometry experiments to show that the inconsistencies between Hall and optical measurements can be largely eliminated, leading to an improved determination of carrier concentrations that is free of systematic errors. We combine these "self-consistent" values with resistivity measurements to extract the mobility of P-, As-, and Sb-doped Ge, and we propose empirical expressions that can be used for modeling the electrical properties of doped Ge materials.

## II. EXPERIMENTAL DETAILS AND STATEMENT OF THE PROBLEM

We performed Hall effect and spectroscopic ellipsometry studies on a set of 40 *n*-type Ge samples doped with P, As, and Sb with carrier concentrations ranging from $n = 7 \times 10^{18}$ cm$^{-3}$ to $n = 1.3 \times 10^{20}$ cm$^{-3}$. The samples were grown as described in Refs. 11,12,17. Briefly, undoped Ge buffer layers were first deposited on (001) Si substrates in a gas-source molecular epitaxy (GSME) chamber using tetragermane Ge$_4$H$_{10}$ as the Ge source. The growth was carried out at temperatures near 350 °C and the samples were typically subjected to *in situ* annealings at temperatures about 650 °C to reduce dislocation densities. The use of higher-order germanes such as Ge$_3$H$_8$ or Ge$_4$H$_{10}$ provides an alternative to the standard two-step approach for the growth of Ge films on Si substrates.[18] The advantage of the polygermane method is that a high-defect sublayer is avoided. The Ge-buffered samples are transferred to an ultra-high vacuum chemical vapor deposition reactor (UHV-CVD), in which mixtures of Ge$_3$H$_8$, H$_2$, and the above-mentioned dopants were used to grow the thin doped films.

The carrier concentrations were obtained from Hall effect measurements and spectroscopic ellipsometry. The Hall measurements were performed at room temperature on approximately 10 mm square samples using a Ecopia 3000 system at room temperature. The magnetic field was measured to be of the form $B = B_0 - Ar^2$, where $r$ is the radial distance to the magnet's axis, with $B_0 = 0.584$ T and $A = 3 \times 10^{-3}$ T/mm$^2$. The field was averaged over the sample's area for carrier concentration determinations. In-Sn contacts were formed at the sample corners for measurements in the van der Pauw configuration. Resistivity measurements were also made with the same arrangement. The contact-size was less than 7% of the sample size, which implies a negligible error for the resistivity and less than 5% error for the Hall coefficient $R_H$. Based on the mobility $\mu$ values shown below, we have $\mu B \ll 1$ for all samples, which means that we can use low-field approximations. Under these conditions the Hall coefficient is given by

$$R_H = -\frac{\gamma_H}{en}, \qquad (1)$$

where $\gamma_H$ is the so-called Hall factor and $e$ is the absolute value of the electron charge.

Infrared spectroscopic ellipsometry measurements were performed on a J. A. Woollam IR-VASE within an energy range 0.03 eV < $E$ < 0.8 eV, using a step size of 1 meV and an angle of 70°. The experimental complex dielectric function below 0.6 eV is assumed to be of the form

$$\hat{\epsilon}(\omega) = \epsilon_{\text{opt}}(\omega) - \frac{1}{\epsilon_0 \rho(\tau \omega^2 + i\omega)} \qquad (2)$$

where $\rho$ is the resistivity and $\tau$ an average relaxation time. These two parameters can be extracted from simultaneous fits of the real and imaginary parts. The function $\epsilon_{\text{opt}}(E)$ is the low-energy extrapolation of the real part of the optical dielectric function. It originates from valence-conduction interband transitions, and it is well approximated by expressions of the form $\epsilon_{\text{opt}}(E) = \epsilon_\infty + \sum_i A_i/(E^2 - E_i^2)$, where the sum is over a series of "poles" related to critical points in the visible dielectric function. The parameters in these expressions can be fit to our own ellipsometric measurements in the visible or taken from the literature. The specific model chosen has a very minor impact on the resistivity and relaxation time extracted from the infrared fits, with differences that never exceed 1%. For a cubic crystal,

$$\frac{1}{\rho\tau} = -\frac{e^2}{4\pi^3\hbar^2}\int_0^\infty dE \frac{\partial f}{\partial E}\int d\mathbf{k}\ \left(\frac{\partial E(\mathbf{k})}{\partial k_x}\right)^2 \delta[E - E(\mathbf{k})] \tag{3}$$

where $f(E)$ is the Fermi-Dirac distribution function and $x$ is a direction along any one of the cubic axes. If the dispersion is parabolic, this expression reduces to

$$\frac{1}{\rho\tau} = \frac{e^2 n}{m^*} \tag{4}$$

where $m^*$ is the conductivity effective mass. For electrons in the conduction band lowest valley, located in Ge around the $L$ point of the Brillouin Zone, this conductivity effective mass satisfies

$$\frac{3}{m^*} = \left(\frac{1}{m_\parallel} + \frac{2}{m_\perp}\right) \tag{5}$$

where $m_\parallel$ is the dispersion mass along the <111> direction and $m_\perp$ the equivalent mass in a perpendicular direction. Eq. (4) can then be used to determine the carrier concentration from the ellipsometry parameters if $m_\parallel$ and $m_\perp$ are known.

For carrier concentration determinations, it is customary to assume a Hall factor $\gamma_H = 1$ and the conductivity effective mass that results from cyclotron resonance, magnetoabsorption, and magnetopiezo-transmission measurements[19-21] of $m_\parallel$ and $m_\perp$, $m^* = 0.12\ m_0$, where $m_0$ is the free electron's rest mass. Using these assumptions, we show in Fig 1(a) the carrier concentrations $n_\text{ellip}$ obtained from ellipsometry meausurements versus the carrier concentrations $n_\text{Hall}$ obtained from Hall measurements. A fit of the form $n_\text{ellip} = a n_\text{Hall}$ gives $a = 0.80\pm 0.02$, indicating a significant systematic error, most likely associated with our choice of the Hall factor and/or the value of the conductivity effective mass. To address these possibilities, we present below a calculation of the effective mass and the Hall factor.

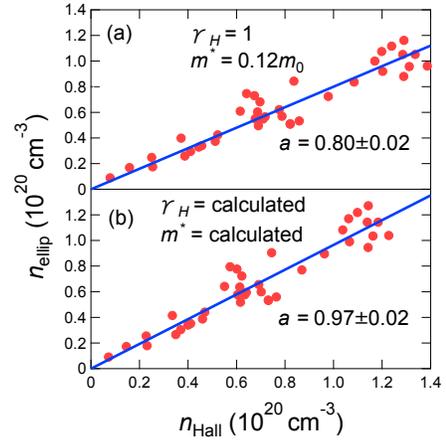

**Figure 1** (a) Carrier concentrations $n_\text{ellip}$ in $n$-type Ge measured by spectroscopic ellipsometry versus carrier concentrations $n_\text{Hall}$ from Hall measurements. A fixed conductivity effective mass $m^* = 0.12 m_0$ was assumed for the ellipsometry analysis, and a Hall factor $\gamma_H = 1$ was assumed for the Hall data. The solid line is a fit $n_\text{ellip} = a\ n_\text{Hall}$.
(b) Same as (a) but after recomputing $n_\text{ellip}$ and $n_\text{Hall}$ using the effective mass from Fig. 2 and the Hall factor from Fig. 3.

### III. EFFECTIVE MASS CALCULATION

The validity of Eq. (4) rests on the assumption of parabolic dispersion. To include non-parabolicity effects, we note that the perpendicular effective mass that appears in Eq. (5) is given in $\mathbf{k}\cdot\mathbf{p}$ theory by[13]

$$\frac{m_0}{m_\perp} = 1 + \frac{\bar{P}^2}{m_0}\left(\frac{1}{E_1} + \frac{1}{E_1 + \Delta_1}\right) \tag{6}$$

where $E_1$ (2.11 eV) and $E_1 + \Delta_1$ (2.31 eV) are the lowest direct band gaps at the $L$ point, and $\bar{P}^2$ is a momentum matrix element that can be determined from the experimental value of $m_\perp$ at 4K.[19] In the limit $\Delta_1 = 0$, the 3×3 $\bm{k} \cdot \bm{p}$ problem leading to Eq. (6) can be trivially diagonalized, leading to a dispersion relation of the form

$$E(\bm{k}) = \alpha_L k_\parallel^2 + \sqrt{\Delta_L^2 + \beta_L^2 k_\perp^2} - \Delta_L \tag{7}$$

where $\alpha_L = \hbar^2/(2m_\parallel)$, $\Delta_L = E_1/2$, and $\beta_L^2 = \hbar^2(\Delta_L + \bar{P}^2/m_0)/m_0$. This expression neglects non-parabolicity in the parallel direction, which is justified by the fact that in this direction $\bm{k} \cdot \bm{p}$ coupling occurs with more distant bands. While derived for the special case $\Delta_1 = 0$, the fact that $\Delta_1$ is considerably smaller than $E_1$ suggests that we use Eq. (7) as an approximate expression in the general case, but redefining the parameter $\Delta_L$ as

$$\frac{p}{\Delta_L} = \frac{1}{E_1} + \frac{1}{E_1 + \Delta_1}, \tag{8}$$

where $p$ is a constant (expected to be close to unity if our approach is reasonable) that can be adjusted to match the dispersion relation obtained from an accurate band structure calculation. We have done this using the 30-band $\bm{k} \cdot \bm{p}$ model from Ref. 22, and we find $p = 0.786$.

The density of states corresponding to Eq. (7) can be written, accounting for the 8 degenerate <111> directions, as

$$g(E) = \frac{4\sqrt{2}}{\pi^2} \left(\frac{m_L}{\hbar^2}\right)^{3/2} \left(E^{1/2} + \frac{2}{3}\frac{E^{3/2}}{\Delta_L}\right) \tag{9}$$

where $m_L = (m_\parallel m_\perp^2)^{1/3}$ is the density-of-states effective mass. If we use Eqs. (7) and (9) in Eq. (3)(3), we find that Eq. (4) can be used if we define the effective mass as

$$\frac{3}{m^*} = \left[\frac{2}{m_\perp} + \frac{1}{m_\parallel}\right] \frac{\mathcal{F}_{1/2}(E_F/k_BT) - \left(\frac{k_BT}{\Delta_L}\right)\left(\frac{2m_\parallel - m_\perp}{2m_\parallel + m_\perp}\right)\mathcal{F}_{3/2}(E_F/k_BT)}{\mathcal{F}_{1/2}(E_F/k_BT) + \left(\frac{k_BT}{\Delta_L}\right)\mathcal{F}_{3/2}(E_F/k_BT)} \tag{10}$$

Here $k_B$ is Boltzmann's constant, $T$ the absolute temperature, and $E_F$ the Fermi level measured from the bottom of the conduction band. The functions $\mathcal{F}_{1/2}(y)$ and $\mathcal{F}_{3/2}(y)$ are Fermi-Dirac integrals defined as in Ref. 23. It is apparent that the effective mass so defined depends on the amount of doping, which affects the position of the Fermi level. Note that the expression that appears in Refs. 13,24 is an expansion of Eq. (10) valid in the limit of small non-parabolicity. We use the exact expression here because such expansion is not accurate enough for our purposes. We computed $m^*$ from Eq. (10) using a Fermi level calculated from the model presented in detail in Ref. 25, which includes the three lowest valleys in the conduction band ($L$, $\Gamma$, and $\Delta$) (with non-parabolicity corrections for the lowest two), and also accounts for non-parabolicity and warping in the valence bands. The model has been shown to reproduce the experimental intrinsic carrier concentrations extremely well.[26] The doping-induced renormalization of all relevant band structure features was included based on results from Ref. 13. To calculate the effective mass at 295 K, we assume that $m_\parallel$ is independent of temperature, so that we can use the 4K value.[19] For $m_\perp$, we can obtain the room temperature value from Eq. (6) using the known temperature dependence of $E_1$ and $E_1 + \Delta_1$ and the expected dependence of $\bar{P}^2$ on the lattice parameter.[13] Results are shown in Fig. 2, and we see that the conductivity effective mass increases rapidly when the doping level moves beyond $10^{19}$ cm$^{-3}$.

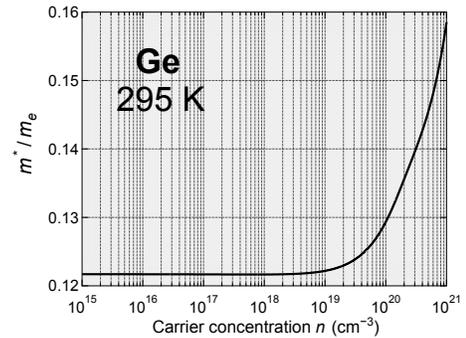

**Figure 2** Conductivity effective mass for $n$-type Ge, calculated using Eq. (10) and experimental values for $m_\parallel$ and $m_\perp$.

## IV. HALL FACTOR CALCULATION

The Hall factor corresponding to electrons in the lowest conduction band valley in Ge is given by[27,28]

$$\gamma_H = \frac{3\left(\frac{\langle \tau_\perp^2 \rangle}{m_\perp^2} + 2\frac{\langle \tau_\perp \tau_\parallel \rangle}{m_\perp m_\parallel}\right)}{\left[2\frac{\langle \tau_\perp \rangle}{m_\perp} + \frac{\langle \tau_\parallel \rangle}{m_\parallel}\right]^2} \qquad (11)$$

where $\tau_\perp$ and $\tau_\parallel$ are the perpendicular and parallel and components of the anisotropic relaxation time tensor. The averages in the expression are defined as

$$\langle h \rangle = \frac{2}{3k_B T} \frac{\int dE\ E^{3/2} h(E) f(E)[1-f(E)]}{\int dE\ E^{1/2} f(E)} \qquad (12)$$

for any function $h(E)$. In the non-degenerate Maxwell-Boltzmann limit, Eq (12) approaches the standard expressions for these averages given in textbooks.[29] For our highly doped samples, however, it is important to account for degeneracy.

In doped Ge, the relaxation time is mainly due to the electron-phonon interaction and to the interaction of carriers with ionized impurities. If we assume that the anisotropy of the relaxation time is only due to ionized impurity scattering, we can write[30]

$$\begin{aligned} \frac{1}{\langle \tau_\parallel \rangle} &= \frac{1}{\langle \tau_{ph} \rangle} + \frac{1}{\langle \tau_{i,\parallel} \rangle} \\ \frac{1}{\langle \tau_\perp \rangle} &= \frac{1}{\langle \tau_{ph} \rangle} + \frac{1}{\langle \tau_{i,\perp} \rangle} \end{aligned} \qquad (13)$$

The mobility is then

$$\mu = \frac{e}{3}\left(\frac{2\langle \tau_\perp \rangle}{m_\perp} + \frac{\langle \tau_\parallel \rangle}{m_\parallel}\right) \qquad (14)$$

We model the lattice scattering by assuming $\tau_{ph} = \tau_{ph} = \frac{a}{k_B T} E^{-\alpha}$, and we adjust the constants $a$ and $\alpha$ to match the experimental temperature dependence of the electron mobility in undoped Ge (Ref. 31). We find $\alpha = 0.69$ and $a = 8.51 \times 10^{-36}$ erg$^{1.69}$ s. For the impurity scattering, we reproduce the calculation of Ito,[30] except that we compute the averages as prescribed in Eq. (12) rather than approximating them by their value at the Fermi level.

The resulting Hall factor is shown in Fig. 3. For self-consistency, the calculated average relaxation times should reproduce the experimental mobility. This is shown in Figure 4 below, and it is apparent that the agreement is excellent. We see that the predicted Hall factor has a complicated dependence on the doping concentration. To understand this behavior, we note that the Hall factor associated with lattice scattering alone is smaller than the Hall factor associated with ionized impurity scattering as shown in elementary calculations of these effects.[29] At low carrier concentrations the Hall factor is dominated by lattice scattering. If we expand Eq. (11) as a series in the small quantities $\tau_{ph}/\tau_{i,\parallel}$ and $\tau_{ph}/\tau_{i,\perp}$, we find that the first-order correction depends on the anisotropy of the ionized impurity scattering, and it is negative when $\tau_{i,\parallel} > \tau_{i,\perp}$, as is the case here. This explains the initial decrease in $\gamma_H$ for concentrations between $n = 10^{15}$ cm$^{-3}$ and $n = 10^{17}$ cm$^{-3}$. For higher concentrations, ionized impurity scattering becomes dominant, and the Hall factor rises, as expected from the elementary results. However, our calculations also show that at as the carrier concentration increases, both Hall factors associated with lattice and ionized impurity scattering decrease as a result of band structure and degeneracy effects, so that a new local maximum is predicted near $n = 10^{19}$ cm$^{-3}$.

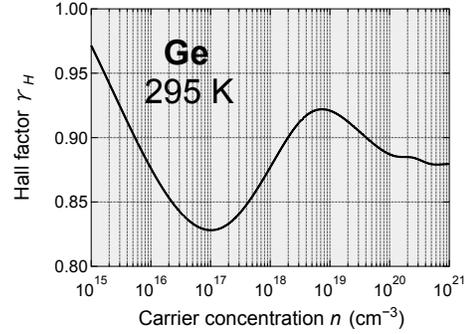

**Figure 3** Hall factor for electrons in $n$-type Ge, calculated as explained in the text.

Comparing Fig. 2 and Fig. 3, we conclude that for concentrations below $10^{19}$ cm$^{-3}$ optical measurements may be less prone to error, since the effective mass is essentially constant while the Hall factor has a strong oscillation. The opposite is true for concentrations above $10^{19}$ cm$^{-3}$, since the effective mass is predicted to change by more than 30% between $n = 10^{19}$ cm$^{-3}$ and $n = 10^{21}$ cm$^{-3}$, while the Hall factor only changes by 4% over the same range.

## V. RECALCULATED CARRIER CONCENTRATIONS

Using the carrier-concentration dependence of the effective masses in Fig. 2 and the carrier concentration of the Hall factor in Fig. 3, we have recomputed the carrier concentrations determined from ellipsometry and Hall experiments. The results are shown in Fig. 1(b), and we see that slope of the $n_{\text{ellip}} = an_{\text{Hall}}$ curve is now $a = 0.97 \pm 0.02$, much closer to unity. We then believe that we have addressed the main contributions to systematic errors in the determination of carrier concentrations.

Further improvements may be possible from the observation that the correction that we applied to the ellipsometry result is not fully consistent with the correction calculated for the Hall results. For the effective mass calculation, we account for non-parabolicity in the context of an isotropic, energy-independent relaxation time. On the other hand, for the Hall factor calculation we consider the full energy dependence and anisotropy of the relaxation time, but assume parabolic dispersion. The inconsistency is mitigated by the fact that at the plasma frequency $\omega_p$ [$\omega_p^2 = ne^2/(m^*\epsilon)$] most relevant for the ellipsometric fits, we find $\omega_p\tau \gg 2$, so that the dielectric function is not too far from the high-frequency limit in which it becomes independent of the relaxation time. On the other hand, since non-parabolicity would affect both the numerator and denominator in Eq. (12), we would expect a partial cancellation, so that the Hall factors should not be strongly affected by non-parabolicity. We note that in order to include non parabolicity in the Hall factor calculation it is not enough to modify Eq. (12) by replacing factors of $E^{1/2}$ by factors of $E^{1/2} + \frac{2}{3}E^{3/2}/\Delta_L$, as suggested by Eq. (9). This is because the additional factor of $E$ in the numerator of Eq. (12), which arises from the electron velocities, is only obtained for parabolic dispersion. Furthermore, the expressions for ionized impurity scattering in Ref. 30 were also derived assuming parabolic dispersion, and they should be modified for full consistency.

Even if the effective mass and Hall factor calculation are made fully consistent, the fact remains that for carrier concentrations in the range shown in Fig. 1, the impurity band and the conduction band have merged.[14] In this regime one may be able to continue using effective masses in the context of a "virtual crystal" description of the electronic structure, but it is by no means obvious that the masses should be the ones computed here. From this perspective, the agreement between Hall and ellipsometry measurements in Fig 1(b) is quite remarkable.

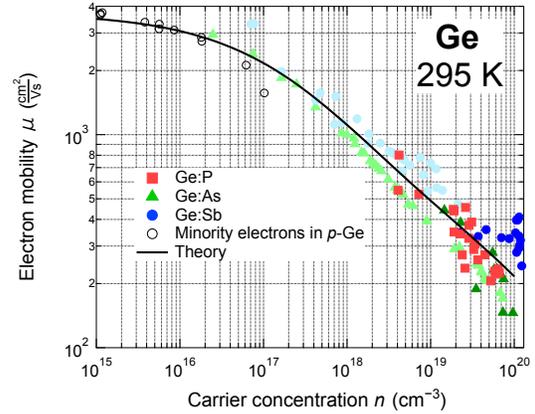

**Figure 4** Electron mobilities in *n*-type Ge at room temperature. Darker colors represent our samples, lighther colors were taken from the literature and corrected using our calculated Hall factors. The empty circles correspond to minority electron mobilities in *p*-Ge at the doping concentrations corresponding to the horizontal axis. The solid line is the theoretical calculation of the mobility, as described in the text. Notice that the excellent agreement with experiment is obtained without any adjustable parameter for the ionized impurity scattering contribution to the mobility.

**VI. ELECTRON MOBILITIES**

Using the recalculated carrier concentrations and the measured resistivities, we have computed the mobilities for all of our samples. The results are shown in Fig. 4, where we have added older data in the literature for As- and Sb-doped Ge.[7,32,33] These earlier papers used $\gamma_H = 1$, and we have corrected their data based on Fig. 3. The results clearly display the well-known Sb > P > As donor-dependence of the mobilities. The theoretical calculation, which assumes exactly the same potential for all donors, runs nicely through the average of the mobilities.

For modeling purposes, it is convenient to have empirical expressions that reproduce the experimental mobilities. Inspired by the work of Hillsum (Ref. 34), we have fit the data with

$$\mu = \frac{\mu_0}{1 + \left(\frac{n}{A \times 10^{17} \text{cm}^{-3}}\right)^{\alpha_0}} \quad \text{for } n \leq 10^{17} \text{cm}^3$$

$$\mu = \frac{\mu_0}{1 + \left(1 - \frac{\alpha_0}{\alpha_1}\right)\left(\frac{1}{A}\right)^{\alpha_1} + \left(\frac{\alpha_0}{\alpha_1}\right)\left(\frac{1}{A}\right)^{\alpha_0 - \alpha_1}\left(\frac{n}{A \times 10^{17} \text{cm}^{-3}}\right)^{\alpha_1}} \quad \text{for } n > 10^{17} \text{cm}^3 \quad (15)$$

The parameters $\mu_0$, $A$, and $\alpha_0$ are constrained to be the same for all donors, and only the parameter $\alpha_1$ is allowed to vary according to donor. The expression in the denominator for $n > 10^{17}$ cm$^{-3}$ ensures that the function $\mu(n)$ and its derivative are continuous at $n = 10^{17}$ cm$^{-3}$. Eq (15) is the simplest way to account for the difference between donors with a single parameter, while ignoring the difference between donors at low concentrations. The fit parameters are shown in Table 1.

|  | $\mu_0$ (cm$^2$/Vs) | $A$ | $\alpha_0$ | $\alpha_1$ |
|---|---|---|---|---|
| Ge:P | 4.90×10$^7$ $T^{-1.66}$ ($T$ in K) | 1.393 | 0.695 | 0.357 |
| Ge:As |  |  |  | 0.424 |
| Ge:Sb |  |  |  | 0.232 |

**Table 1** Parameters of fits of the electron mobility in *n*-Ge using the model expression in Eq. (15)

## VII. MAJORITY VS MINORITY MOBILITIES

We have added to Fig. 4 the minority electron mobilities in *p*-type Ge as measured by Prince.[35] We see that within the error of the data the mobility values for majority and minority electrons are very similar and there is little motivation to model them separately for device simulations. This result is consistent with prior observations in silicon,[36,37] although the latter are based on modern minority mobility measurements that may be more reliable.

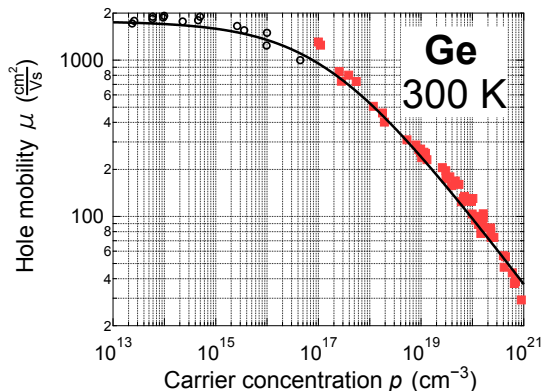

**Figure 5** Hole mobilities in *p*-type Ge at room temperature. Red squares are data from Trumbore (Ref. 38). The empty circles correspond to minority hole mobilities in *n*-Ge at the doping concentrations corresponding to the horizontal axis. The solid line is a fit with Eq. (15).

As to the mobilities of minority holes, a comparison with majority holes would require a calculation of the Hall factor for holes similar to that performed for electrons. Unfortunately, this is a more challenging task given the strong non-parabolicity and warping of the valence bands and the presence of interband scattering. Current available data for majority holes[38] are shown in Fig. 5. When the measured mobility from minority holes[35] in *n*-type Ge are added to the figure, no clear indication of a mismatch is observed. The solid line shows a fit of the majority hole data with Eq. (15). The parameter $\mu_0$ is set to $\mu_0 = 1776$ cm$^2$/(Vs) to match the hole mobility in intrinsic Ge(Ref. 39), and constraining $\alpha_0 = \alpha_1$ produces a very good fit with $A = 1.393$, and $\alpha_0 = 0.434$.

## CONCLUSIONS

We have presented a methodology to reconcile carrier concentrations in *n*-type Ge as determined from Hall and ellipsometry measurements. For the optical measurements, we computed the doping dependence of the effective conductivity mass assuming a non-parabolicity model that was fit to accurate band structure calculations. For the correction of the Hall measurements, we computed the Hall factor assuming lattice and ionized impurity scattering in a model. The model is fit to the electron mobility in undoped Ge and uses no additional adjustable parameters to account for ionized impurity scattering in *n*-type samples. It gives excellent agreement with the experimental mobility, which suggest it is appropriate for calculating Hall factors.

The corrected carrier concentrations were combined with experimental resistivities to compute carrier mobilities, and these mobilities were fit with empirical expressions that can be used for device modeling. A comparison of minority and majority carrier mobilities does not reveal any significant difference for the data currently available for Ge.

## ACKNOWLEDGEMENT

This work was partially supported by the US National Science Foundation under grant DMR-2119583.